\documentclass[twocolumn, journal]{IEEEtran}
 
\usepackage{booktabs} % For professional looking tables
\usepackage{multirow}
\usepackage{mathtools}

\usepackage{circuitikz}

\usepackage{nccmath}

\usepackage{amssymb}
 
\usepackage{enumitem}
\usepackage{stackengine}

\usepackage{smartdiagram}
\usesmartdiagramlibrary{additions}

\graphicspath{{Figs/}}

\usepackage{pdfpages}
\usepackage{pdflscape}
\usepackage{amsmath}
\usepackage{amsthm}

\usepackage{amsfonts}

\usepackage{color}

\usepackage{setspace}
\usepackage{enumitem}
\usepackage{xcolor}

\usepackage{bm}
\usepackage{bbm}

\usepackage{tikz}
\usepackage{pgf}
\usetikzlibrary{arrows,automata}
\usetikzlibrary{shapes}
\tikzstyle{data44}=[rectangle split,rectangle split parts=2,draw,text centered]

\usepackage{subfigure}

\DeclareMathAlphabet{\mathcal}{OMS}{cmsy}{m}{n}

\usetikzlibrary{matrix}

\tikzset{
  BarreStyle/.style =   {opacity=.3,line width=14 mm,color=#1},
  node style ge/.style={},
  node style sp/.style={},
  yl/.style={},
  arrow style mul/.style={},
}

\newtheoremstyle{mytheoremstyle} % name
        {\topsep}                    % Space above
        {\topsep}                    % Space below
        {\fontfamily{ptm}\selectfont}                   % Body font
        {}                           % Indent amount
        {\itshape\fontfamily{ptm}\selectfont}                   % Theorem head font
        {:}                          % Punctuation after theorem head
        {.5em}                       % Space after theorem head
        {}  % Theorem head spec (can be left empty, meaning ‘normal’)
\theoremstyle{mytheoremstyle}

\newtheorem{definition}{Definition}

\usepackage{threeparttable,booktabs}

\usepackage[colorlinks,
            linkcolor=blue,
            anchorcolor=blue,
            citecolor=blue
            ]{hyperref}

\usepackage{cite}

\usepackage{mdwlist}

\usepackage[linesnumbered,ruled,vlined]{algorithm2e}
\SetKwInput{KwInput}{Input}                % Set the Input
\SetKwInput{KwOutput}{Output}   
\let\oldnl\nl% Store \nl in \oldnl
\newcommand{\nonl}{\renewcommand{\nl}{\let\nl\oldnl}}% Remove line number for one line

\usepackage{etoolbox}

\makeatletter
\patchcmd{\@algocf@start}% <cmd>
  {-1.5em}% <search>
  {0pt}% <replace>
  {}{}% <success><failure>
\makeatother

\usepackage{textcomp}

\usepackage{stmaryrd}

\usepackage{comment}

\usepackage{graphicx,subfigure}

\usepackage{stfloats}
\usepackage{tabularx }

\usepackage{makecell}

\usepackage{colortbl}

\newcolumntype{P}[1]{>{\centering\arraybackslash}p{0.66cm}}
 
\usepackage{anyfontsize}
\usepackage{t1enc}

\newcommand{\T}{^{\scriptscriptstyle\rm T}}
\usepackage[strict]{changepage}
\usepackage{relsize}

\usepackage{diagbox}

\usepackage{geometry}
\begin{document}

% \title{\Large{Neural-Analytica Energy Functions for Power System Stability Analysis}}

\title{   
    \vspace{-12pt}
    \begingroup
    \fontsize{18pt}{8pt}\selectfont
{Neural-Analytic Energy Functions for Power System Stability Analysis}
\endgroup}

\author{Tong Han,~\IEEEmembership{Member, IEEE}, 
        Yan Xu,~\IEEEmembership{Senior Member, IEEE},
        Rui Zhang,~\IEEEmembership{Member, IEEE}
% \thanks{This work was supported by ...\color{white}{the Research Grants Council of the Hong Kong Special Administrative Region through the Theme-Based Research Scheme under Project T23-701/14-N.}}
\thanks{The authors are with the School of Electrical and Electronic Engineering, Nanyang Technological University, Singapore 639798. \\
This work was originally submitted to \textit{IEEE Power Engineering Letters} but was not accepted, mainly because its content better suited a full paper. It has since been extended and combined with another work into a full paper, accepted for publication in \textit{IEEE Transactions on Power Systems}~\cite{han-temp}.}

\vspace{-20pt}

}

% modify

\maketitle

\begin{abstract} 
    This letter proposes a novel form of energy function (EF), called neural-analytic EF, for the stability analysis of power systems. 
    The neural-analytic EF is designed as the sum of neural networks (NNs)-based components and an analytic EF, with the structure informed by the existing EFs derived analytically. It thus integrates NN's expressivity with the generalizability and scalability of analytic EFs. 
    The satisfactory neural-analytic EF is further derived by iterative NN training with a tailored loss function. 
    Numerical results finally demonstrate its superiority. 
\end{abstract}

\vspace{-4pt}

\IEEEpeerreviewmaketitle

\vspace*{-4pt}
\section{Introduction} 
\vspace{-2pt}

The construction of energy functions (EFs) is the groundwork of direct methods for the stability analysis of power systems \cite{4-49}. 
Initially, EFs were constructed with only generators' second-order dynamics   \cite{4-1701, 4-1863}. 
After developing the structure-preserving EFs with also load dynamics \cite{4-999-102}, more practical EFs were later constructed \cite{4-49, 4-50}. 
These existing EFs are derived analytically by transforming system dynamic models into a specific form (e.g., the Lure form \cite{4-332}). 
However, these analytic EFs are known to produce overly conservative outcomes in stability analysis \cite{4-1058}. 
Recently, the neural Lyapunov method has been introduced to power systems \cite{4-1377, 4-1650}, utilizing neural networks (NNs) to construct Lyapunov functions (LFs). 
While the neural LFs offer less conservative stability results, they pose challenges not only in adapting to changes in system structure, but also in being constructed for large-scale systems due to their inherent lack of scalability. 
In this letter, inspired by the neural Lyapunov method, we propose the neural-analytic EF used for stability analysis. It integrates the expressivity of NNs with the generalizability and scalability of analytic EFs. 
A training method is also devised to ensure validity and reduce conservatism of the neural-analytic EF.

\vspace{-4pt}
\section{Dynamic Model and Energy Function}
\vspace{-2pt}

The proposed method is presented using a simplified transmission system model. 
Generators are represented by the classical model with perfect voltage control. Loads are modelled as a hybrid of constant power and frequency-dependent loads, and losses are ignored. 
Notably, the proposed method remains applicable to more complex system dynamics. 
It is further assumed that the system consists of $n_{\rm a}$ buses, including $n_{\rm g}$ generator buses and $n_{\rm d}$ load buses, and $n_{\rm b}$ lines. Denote by $\mathcal{V}$ the bus set, and $\mathcal{E}$ the line set. 
The generator (or load) at bus $i \in \mathcal{V}$ is referred to as generator (or load) $i$. The phase angle of an arbitrary load bus $i^*$ serves as the angle reference. 
Then all buses are classified into three types based on their dynamics: type 1 buses including all generator buses, type 2 buses including all load buses except for bus $i^*$, and type 3 buses including bus $i^*$. 
Let $\alpha(i)$ be the type of bus $i$, $\mathcal{A} \!\!=\!\! \{1, 2, 3\}$ the set of bus types, and $\mathcal{V}_i$ the set of buses of type $i$. 
For any bus pair $(i, j)$ with $i, j \!\in\! \mathcal{V}$ and $i \neq j$, its type, denoted as $\beta(i, j)$, is defined as $\alpha(i)$-$\alpha(j)$. 
Let $\mathcal{B} \!=\! \{1\text{-}1, 2\text{-}2, 1\text{-}2, 1\text{-}3, 2\text{-}3\}$ collect all possible types of bus pairs. 
The dynamics for each type of buses form the state-space system model as follows: 
\vspace*{-7pt}
\begin{equation}\label{eq-8-2-1}
    \vspace*{-4pt}
    \begin{aligned}
        &\!\!
        \left.   
            \begin{aligned}
                & \dot{\delta_i} \!=\! \omega_i \!-\! \omega_{i^*}, m_i \dot{\omega}_i \!=\! p_{{\rm g}, i}  \!-\! d_{{\rm g}, i} \omega_i \!-\! b_i \sin(\delta_i \!-\! \theta_i) \\[-0.7mm]
                & \epsilon \dot{\theta_i} = - b_i \sin(\theta_i - \delta_i) - \textstyle \sum\nolimits_{j \in \mathcal{C}_i}  b_{ij} \sin(\theta_i - \theta_j)
            \end{aligned}
            \right] i \!\in\! \mathcal{V}_1
        \\[-1.4mm]
        & d_{{\rm d},i} \dot{\theta}_i   \!=\!  \!- p_{\rm d, i} \!-\!  \textstyle {\sum\nolimits_{j \in \mathcal{C}_i}} b_{ij} \sin(\theta_i - \theta_j) \!-\! d_{{\rm d},i} \omega_{i^*} ~~~ i \in \mathcal{V}_2
    \end{aligned}
\end{equation}
where  $
    \omega_{i^*} \!\!\doteq\!\! (1 / d_{{\rm d},i^*}) \!\cdot\! (- p_{{\rm d}, i^*} \!-\!\! \textstyle {\sum\nolimits_{j \in \mathcal{C}_{i^*}}} \!\! b_{i^*j} \sin(\theta_{i^*} \!-\! \theta_j))$ and $ \theta_{i^*} \!\doteq\! 0$; 
$\theta_i$ is the voltage angle of bus $i$; 
$\delta_i$ is the rotor angle, 
$\omega_i$ is the rotor angle speed, 
and $p_{{\rm g}, i}$ is the mechanical input power, all of generator $i$; 
$p_{{\rm d}, i}$ is the constant load power, and $d_{{\rm d}, i}$ is the frequency coefficient, both of load $i$; 
$b_i \doteq {e_i u_i}/{x_i}$ with $x_i$ being the transient reactance and $e_i$ the electromotive force of generator $i$, and $u_i$ being the voltage magnitude of bus $i$; 
$b_{ij} \!\doteq\! {u_i u_j}/{x_{ij}}$ with $x_{ij}$ being the reactance of line $ij$; 
$m_i$ and $d_{{\rm g}, i}$ are respectively the inertia and damping coefficients of generator $i$, 
$\mathcal{A}_i$ is the set of buses connecting with bus $i$, 
$\epsilon$ is a sufficiently small positive numbers introduced by the singular-perturbation approach \cite{4-49}. 
Furthermore, the state-space system dynamic model can be written in a compact form as follows: 
\vspace*{-4pt}
\begin{equation}\label{eq-8-2-2}
    \vspace*{-4pt}
    \dot{\bm{x}} = f_{k}(\bm{x}; \bm{\rho})
\end{equation}
where $k \!\in\! \mathbb{K}$ with $\mathbb{K}$ being its domain, represents the system structure that comprises the number and type of buses, and network topology; 
$\bm{x} \!\in\! \mathbb{R}^n$ with $n \!=\! 3 n_{\rm g} \!+\! n_{\rm d} \!-\! 1$, is the state vector formed by $\theta_i$ of all buses except for $i^*$, and $\delta_i$ and $\omega_i$ of all generators; 
$\bm{\rho} \!\in\! \mathbb{R}^m\!$ with $m \!=\! 2 n_{\rm d} \!+\! 4n_{\rm g} \!+\! n_{\rm b}$ is the parameter vector formed by $m_i$, $d_{{\rm g}, i}$, $p_{{\rm g}, i}$, $b_i$, $d_{{\rm d}, i}$, $p_{{\rm d}, i}$ and $b_{ij}$ of all associated components; 
and $f_k: \mathbb{R}^{n} \mapsto \mathbb{R}^{n}$ is the vector field of the system with structure $k$. 
The EF of system (\ref{eq-8-2-1}) is defined as follows \cite{4-49}: 

\vspace{-0pt}
\begin{definition}[Energy function]\label{def-8-2-1}
    The EF of system (\ref{eq-8-2-1}) is any differentiable function $V_k\!:\! \mathbb{R}^{n + m} \!\mapsto\! \mathbb{R}$ satisfying the following three conditions: 
    (i)  $\mathcal{L}_{f_k(\bm{x}; \bm{\rho} )} V_k(\bm{x}; \bm{\rho}) \!=\! [\frac{ \partial V_k(\bm{x}; \bm{\rho})  }{ \partial \bm{x} }]\T f_k(\bm{x}; \bm{\rho}) \leq 0$;  
    (ii) if $\bm{x}(t)$ is not an equilibrium point of system (\ref{eq-8-2-1}), the set $\{ t | \mathcal{L}_{f_k(\bm{x}; \bm{\rho} )} V_k(\bm{x}(t); \bm{\rho}) \!=\! 0 \}$ has measure 0 in $\mathbb{R}$; and
    (iii) $V_k(\bm{x}(t); \bm{\rho})$ for any given $\bm{\rho}$ is bounded gives $\bm{x}(t)$ is bounded. 
\end{definition}

\vspace{-4pt}
\section{Construction of the Neural-Analytic EF}

\subsection{Designing of the Neural-Analytic EF} 

The existing analytic EFs for system (\ref{eq-8-2-1}) are expressed as the sum of generators' kinetic energy and branches' potential energy. The kinetic energy of a generator depends on its states and parameters, and in some forms (e.g., the EFs in \cite{4-332}), also depends on the states and parameters of the other generators. The potential energy of a branch depends on its parameters and the state variables coupling its two terminals. 
Essentially, the EFs can be viewed as the total energy associated with each bus and each line. 
Also, systems with different structures can share the same EF form. Therefore, we design the \textit{neural-analytic EF} candidate for system (\ref{eq-8-2-1}) in a general form as follows:
\vspace*{-3pt}
\begin{equation}\label{eq-8-2-3} 
    \vspace*{-3pt}
    \!\!\!\!\!
    \begin{aligned}
        & V_{k}(\bm{x}; \bm{\rho} | \bm{\xi} ) \!=\!  \frac{1}{n} \! \big[ \textstyle {\sum\limits_{i \in \mathcal{V}}} \! \textstyle {\sum_{j \in \mathcal{V} \backslash \{i\}  }}   E_{\beta(i, j)} (\bm{x}_i, \bm{x}_j; \bm{\rho}_i,\! \bm{\rho}_j  | \bm{\xi}_{\beta(i, j)} )  \\[-1mm]
        & +  \textstyle {\sum\limits_{i \in \mathcal{V}}}  E_{\alpha(i)}(\bm{x}_i; \bm{\rho}_i | \bm{\xi}_{\alpha(i)} )   \!+\!\!\! \textstyle {\sum\limits_{(i,j) \in \mathcal{E}}}  E_{\rm e} (\bm{\chi}_i, \bm{\chi}_j; \bm{\rho}_{ij} | \bm{\xi}_{\rm e} ) \big] \!+\! V_k^*  
    \end{aligned}
\end{equation}
where $\bm{x}_i$ is the vector formed by the entries of $\bm{x}$ and $\theta_{i^*}$ that are associated with bus $i$, 
with $\bm{x}_i \!=\! [\delta_i, \omega_i, \theta_i]$ if $\alpha(i) \!=\! 1$, 
$\bm{x}_i \!=\! [\theta_i]$ if $\alpha(i) \!=\! 2$, 
and $\bm{x}_i \!=\! [\theta_{i^*}]$ if $\alpha(i) \!=\! 3$; 
$\bm{\rho}_i$ is the vector formed by the entries of $\bm{\rho}$ associated with bus $i$, with 
$\bm{\rho}_i \!=\! [m_i, d_{{\rm g}, i}, p_{{\rm g}, i}, b_i]\T$ if $\alpha(i) \!=\! 1$, 
$\bm{\rho}_i \!=\! [d_{{\rm d}, i}, p_{{\rm d}, i}]\T$ if $\alpha(i) \!=\! 2$ or 3; 
$\bm{\chi}_i \!=\! [\theta_i]$ is the vector formed by the entries of $\bm{x}_i$ and $\theta_{i^*}$ that couple bus $i$ with the other buses; 
$\bm{\rho}_{ij} \!=\! [b_{ij}]$ is the vector formed by the parameters of line $(i, j)$; 
$E_{\alpha(i)}$, $E_{\beta(i, j)}$, and $E_{\rm e}$ denote single-output NNs respectively for buses of type $\alpha(i)$, bus pairs of type $\beta(i, j)$, and lines; 
$\bm{\xi}_{\alpha(i)}$, $\bm{\xi}_{\beta(i, j)}$, and $\bm{\xi}_{\rm e}$ are separately the aggregated parameter vectors of the associated NNs; 
$\bm{\xi}$ is the vector concatenating $\bm{\xi}_l$ for all $l \!\in\! \mathcal{A} \!\cup\! \mathcal{B}$ and $\bm{\xi}_{\rm e}$; 
$V_k^*$ is an arbitrary analytic (accurate or approximated) EF for system (\ref{eq-8-2-1}), e.g., $V_k^* = \frac{1}{n} [ \sum_{i \in \mathcal{V}_1} [ \frac{1}{2} m_{{\rm g}, i} \omega_i^2 - b_i \cos(\delta_i - \theta_i) - p_{{\rm g}, i} \delta_i] \!-\! \sum_{(i,j) \in \mathcal{E}} b_{ij} \cos(\theta_i \!-\! \theta_j) \!-\! \sum_{i \in \mathcal{V}_2} p_{{\rm d}, i} \theta_i ]$ taken from \cite{4-49}. 

% \vspace{-5pt}
\subsection{Training of the Neural-Analytic EF}
% \vspace{-4pt}

Our objective is to construct local EFs within specific domains of $\bm{x}$ and $\bm{\rho}$ for all system structures in $\mathbb{K}$, rather than seeking global EFs. 
Local EFs can be sufficient if the domains cover all common values of system parameters and post-fault state trajectories of interest. 
Let $\mathbb{X}_k$ and $\mathbb{P}_k$ with $k \!\!\in\!\! \mathbb{K}$ be the domains of $\bm{x}$ and $\bm{\rho}$ where the local EFs are valid. 
Then the problem amounts to finding the value of $\bm{\xi}$, i.e., training all the NNs, such that $V_k$ given by (\ref{eq-8-2-3}) satisfies conditions (i) and (ii) in Definition \ref{def-8-2-1} for all $k \!\in\! \mathbb{K}$, $\bm{x} \!\in\! \mathbb{X}_k$ and $\bm{\rho} \!\in\! \mathbb{P}_k$, and the stability results obtained using $V_k$ are minimally conservative. 
Condition (iii) is omitted here since we construct local EFs.

First, consider a sample set of system structures, denoted as $\mathcal{K} \!\subset\! \mathbb{K}$; and a sample set of the pair $(\bm{x}, \bm{\rho})$ for each $k \!\in\! \mathcal{K}$, denoted as $\mathcal{Y}_k \!\subset\! \mathbb{X}_k \!\times\! \mathbb{P}_k$. Also, let $\mathcal{M}$ be a sample set of pair $(k, \bm{\rho})$, with $k \in \mathcal{K}$ and $\bm{\rho}$ taking one value in $\mathcal{Y}_k$ for each $k$. 
For each system with $(k, \bm{\rho}) \in \mathcal{M}$, denote by $\bm{x}_{\rm s}$ a stable equilibrium point of interest, $\mathcal{X}_{\rm u}$ the set of all type-1 unstable equilibrium points (UEPs) whose one-dimensional unstable manifold converges to $\bm{x}_{\rm s}$, 
and $\mathcal{X}_{\rm b}$ a set of system states that are located inside the region of attraction (RoA) of $\bm{x}_{\rm s}$ and near its boundary. 
Note that the superscripts $(k, \bm{\rho})$ for $\bm{x}_{\rm s}$, $\mathcal{X}_{\rm u}$ and $\mathcal{X}_{\rm b}$ are omitted for simplicity. 
Then, 
the value of $\bm{\xi}$ that makes the EF candidate valid w.r.t. the sample sets can be obtained by training all NNs using the following loss function: 
\vspace{-4pt}
\begin{equation}\label{eq-8-2-4} 
    \vspace*{-4pt}
    L(\bm{\xi}) \!=\! (1 \!-\! w) \underbrace{ \frac{1}{ |\mathcal{K}|} \! \medmath{\sum\nolimits_{k \in \mathcal{K}}} \frac{1}{|\mathcal{Y}_k|} L_1^k }_{ \eqqcolon L_1} + w \underbrace{ \frac{1}{|\mathcal{M}|} \medmath{\sum\nolimits_{(k, \bm{\rho}) \in \mathcal{M}} } L_2^{k, \bm{\rho}}  }_{ \eqqcolon L_2 }   
\end{equation}
with 
\vspace*{-4pt}
\begin{equation}\label{eq-8-2-5} 
    \vspace*{-4pt}
    L_1^k \!=\!\!\!\!\!\!\! \medmath{\sum_{(\bm{x}, \bm{\rho}) \in \mathcal{Y}_k }} \!\!\!\!\!\! \max( 0, \mathcal{L}_{f_k} V_k ) \!+\! h_{\phi}(\!  \frac{\Vert f_k \Vert_2}{n}  \!) \cdot \max(0,\!  \frac{\mathcal{L}_{f_k}\! V_k}{n}  + \phi'  )
\end{equation}
\begin{equation}\label{eq-8-2-6} 
    \vspace*{-3pt}
    L_2^{k, \bm{\rho}} =  \frac{1}{ |\mathcal{X}_{\rm b}| } \medmath{\sum\nolimits_{\bm{x} \in \mathcal{X}_{\rm b} }} \Big[ \frac{V_k - \min\nolimits_{\bm{x}_{\rm u} \in \mathcal{X}_{\rm u}} V_{k}(\bm{x}_{\rm u}; \bm{\rho} | \bm{\xi} )}{ V_k - V_k( \bm{x}_{\rm s}; \bm{\rho} | \bm{\xi} ) } \Big]^2
\end{equation}
where $|\cdot|$ is the set cardinality; 
$w$ is the scalar weight; 
$\phi$ and $\phi'$ are small positive numbers; 
$h_{\phi}\!: \!\mathbb{R} \!\mapsto\! \mathbb{R}$ is define as 0 when the input is less than $\phi$, and as 1 otherwise; 
we write $f_k$ and $V_k$ instead of $f_k(\bm{x}; \bm{\rho})$ and $V_k(\bm{x}; \bm{\rho} | \bm{\xi})$ for notational simplicity. 
The term $L_1^k \!\geq\! 0$ measures the violation of condition (i) and (ii) for system structure $k$ over the sample set $\mathcal{Y}_k$;  
$L_2^{k, \bm{\rho}}$ measures, for structure $k$ with parameter $\bm{\rho}$, the average relative deviation of the EF values at the closest UEP and points near the RoA boundary. Smaller $L_2^{k, \bm{\rho}}$ values indicate less conservatism of the RoA yielded by the EF. 
Training the NNs aims to minimize $L_1$ to 0 and reduce $L_2$ as much as possible. The weight $w$ can thus be adjusted using an exponential decay scheme.

After the training, for any $k \in \mathcal{K}$, validity of the obtained EF over the entire subdomains of $\bm{x}$ and $\bm{\rho}$ can be verified in two approaches. 
The first, called empirical verification, is to use sufficient samples from $\mathbb{X}_{k} \times \mathbb{P}_k$. The second one, called formal verification, solves the following satisfiability problem: 
\vspace{-2pt}
\begin{equation}\label{eq-8-2-7} 
    \vspace{-3pt}
    \begin{bmatrix}
        \bm{x} \!\in\! \mathbb{X}_k \land \\[-0.5mm]
        \bm{\rho} \!\in\! \mathbb{P}_k
    \end{bmatrix}
    \land
    \begin{bmatrix}
        ( \mathcal{L}_{f_k} V_k \!\geq\! \varepsilon  \land  \Vert f_k \Vert_2 \!\leq\! n \cdot \phi  ) \lor \\[-0.2mm]
        ( \mathcal{L}_{f_k} V_k \!\geq\! 0 \land \Vert f_k \Vert_2 \!\geq\! n \cdot \phi )
    \end{bmatrix}
\end{equation}
with $\varepsilon > 0$ being the error tolerance. Solving problem (\ref{eq-8-2-7}) can identify counterexamples of $\bm{x}$ and $\bm{\rho}$ where condition (i) or (ii) does not hold with the EF, or encounter infeasibility, thereby verifying the validity of the EF for system structure $k$.

\begin{figure}[t!]
	\centering 
    \includegraphics[scale=0.9]{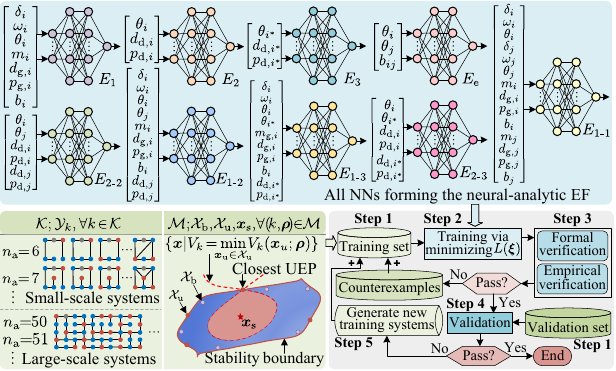}   
    \vspace{-10pt}
	\caption{Illustration and flowchart of the proposed NN training method.}
	\label{fig-8-2-1}
    \vspace{-20pt}
\end{figure}

The above training and verification face two limitations: i) the formal verification that provides rigorous validity guarantee is demanding for large-scale systems, and ii) the obtained EF may not generalize well to changes in system structure. 
These limitations can be further addressed by leveraging the inherent generalizability of EFs. 
First, if the EF has rigorous and empirical validity guarantees for a sufficient number of small and large systems, respectively, the rigorous validity guarantee for the large systems can be almost assured. 
Second, by expanding the system structure set $\mathcal{K}$, the likelihood that the obtained EF is valid for arbitrary system structures also increases. 
Accordingly, we can construct a satisfactory EF by iteratively training the NNs with progressively augmented training data, as depicted in Fig. \ref{fig-8-2-1}, through the following steps: 

\textit{Step 1: Generation of the training and validation sets}. 
The training set includes: system structure set $\mathcal{K}$, set $\mathcal{Y}_k$ for each $k \!\in\! \mathcal{K}$, and sets $\mathcal{X}_{\rm u}$ and $\mathcal{X}_{\rm b}$ and $\bm{x}_{\rm s}$ for each $(k, \bm{\rho}) \!\in\! \mathcal{M}$. 
Set $\mathcal{K}$ consists of $\eta_1 \times \eta_2$ small system structures and $\eta_3$ large system structures. 
For the former, the system size $n_{\rm a} \!=\! 6, 7, ..., 5 \!+\! \eta_1$, and each system size has $\eta_2$ different configurations of network topology and bus types; for the latter, $n_{\rm a} \!=\! 50, 51, ..., 49 \!+\! \eta_3$, and each system size has one configuration. 
% The small and large systems are constructed manually and using SynGrid \cite{4-1866}, respectively, 
Different system structures can be synthesized using SynGrid \cite{4-1866}. 
For each $k \!\in\! \mathcal{K}$, $\eta_4^k$ samples from $\mathbb{X}_k \!\times\! \mathbb{P}_k$ form $\mathcal{Y}_k$. 
Set $\mathcal{M}$ is formed by $\eta_5$ system structures from $\mathcal{K}$ with $n_{\rm a} \!\in\! \{6, 7, 8\}$ and one value of $\bm{\rho}$ in $\mathcal{Y}_k$ for each system structure. 
Set $\mathcal{X}_{\rm b}$ is obtained by time-domain simulations; and set $\mathcal{X}_{\rm u}$ is obtained by solving a series of non-convex optimization problems using Gurobi 11.0. 
The validation set consists of $\eta_6$ small and $\eta_7$ large system structures which are not included in the training set, and $\eta_8$ samples of $(\bm{x}, \bm{\rho})$ for each large system structures.

\textit{Step 2: Training}. Train the NNs via minimizing $L(\bm{\xi})$ using the current training set and the Adam optimizer. The training terminates when $\sum_{k \in \mathcal{K}} L_1^k \!=\! 0$ for consecutive 2 epochs. 

\textit{Step 3: Verification}. Verify the validity of the EF with the trained NNs for all system structures in the training set. Formal and empirical verifications are performed for the small and large system structures, respectively.  
If the verification is passed for all the system structures, proceed to \textit{step 4}; otherwise, add the identified counterexamples and $\eta_9$ neighboring samples for each to the training set, and return to \textit{step 2}. 

\textit{Step 4: Validation}. The validity of the obtained EF w.r.t. an arbitrary system structure is assessed similarly to the verification in step 3 but using the validation set. 
If the validation is passed for all system structures in the validation set, the iteration ends; otherwise, proceed to \textit{step 5}.

\textit{Step 5: Augmentation of the training set.} 
Add new system structures and their associated sample sets $\mathcal{Y}_k$ to the training set, then return to \textit{step 2}. 
Let $n_{\rm as}$ and $n_{\rm al}$ be the largest system size of small and large system structures in the current training set, respectively. 
The added system structures consist of $\eta_2$ small system structures, each with an identical system size of $n_{\rm as}+1$ but different configurations of network topology and bus types; and $\eta_{10}$ large system structures with different system sizes ranging from $n_{\rm al} + 1$ to $n_{\rm al} + \eta_{\rm 10}$. 

\vspace*{-4pt}
\section{Numerical Results}

This section demonstrates the effectiveness and superiority of the neural-analytic EF. 
The parameters are set as follows:  
for all $\mathbb{X}_k$,  
$\theta_i, \delta_i \!\!\!\in\!\! [-2\pi, 2\pi]$, 
$\omega_i \!\!\in\!\!\! [-0.2, 0.2]$;  
for all $\mathbb{P}_k$, 
$p_{{\rm g}, i}, p_{{\rm d}, i}\!\!\in\!\! [0.5, 1.5]$, $\!\sum_{i \in \mathcal{V}_3} \! p_{{\rm d}, i} \!\!-\!\! \sum_{i \in \mathcal{V}_2} \! p_{{\rm g}, i} \!\!\in\!\! [0.5, 1.5]$, 
$b_i, b_{ij}\!\!\in\! [\frac{1}{0.3}, \frac{1}{0.1}]$, 
$m_i \!\!\in\!\! [2, 10]$,  
$d_{{\rm g}, i}\!\!\in\!\! [4, 20]$, 
$d_{{\rm d}, i} \!\!\in\!\! [4, 6]$;  
all values are in p.u. except $\theta_i$ and $\delta_i$; 
all NNs are with the SiLU activation function and 3 hidden layers with widths of 8, 8, and 16 times the input dimension, respectively;  
$\phi \!\!=\!\! 10^{-2}\!$, 
$\phi' \!\!=\!\! 10^{-5}\!$, 
$\varepsilon \!\!=\!\! 10^{-3}$, 
$\eta_1\!\!=\!\!10$, $\eta_2\!\!=\!\!20$, $\eta_3\!\!=\!\!50$, $\eta_4^k\!\!=\!\!2 \!\times\! 10^4$, $\eta_5\!\!=\!\!10$, $|\mathcal{X}_{\rm b}|\!\!=\!\!100$, $\eta_6\!\!=\!\!50$, $\eta_7\!\!=\!\!100$, $\eta_8\!\!=\!\!10^8$, $\eta_9\!\!=\!\!99$, $\eta_{10}\!\!=\!\!20$; 
the batch size is 51,200.

\subsubsection{Training process} 
\begin{figure}[t!]
	\centering 
    \includegraphics[scale=1.0]{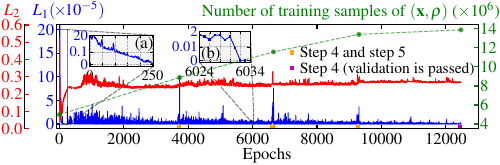}   
    \vspace{-12pt}
	\caption{Curves of $L_1$, $L_2$, and training sample size for $(\bm{x}, \bm{\rho})$  during training.}
	\label{fig-8-2-r3}
    \vspace{-8pt}
\end{figure}

Fig. \ref{fig-8-2-r3} shows the curves of $L_1$, $L_2$, and the training sample size for $(\bm{x}, \bm{\rho})$. 
The initial \textit{training session} (i.e., one execution of step 2) reduces $L_1$ to 0 after approximately 250 epochs, as shown in Fig. \ref{fig-8-2-r3}-(a). 
Differently, each subsequent training session achieves the same minimization of $L_1$ with notably fewer epochs—such as 10 epochs, as shown in Fig. \ref{fig-8-2-r3}-(b). 
The training set is augmented three time by step 5, and the training sample size for $(\bm{x}, \bm{\rho})$ is increased from $5 \!\times\! 10^6$ to nearly $14 \!\times\! 10^6$. 
After about 12,000 epochs, the validation passed in step 4 concludes the entire training, with a reduction of $L_2$ approximately from 0.5 to 0.3.

\subsubsection{Validation on post-fault system states} 
\begin{figure}[t!]
	\centering 
    \includegraphics[scale=0.95]{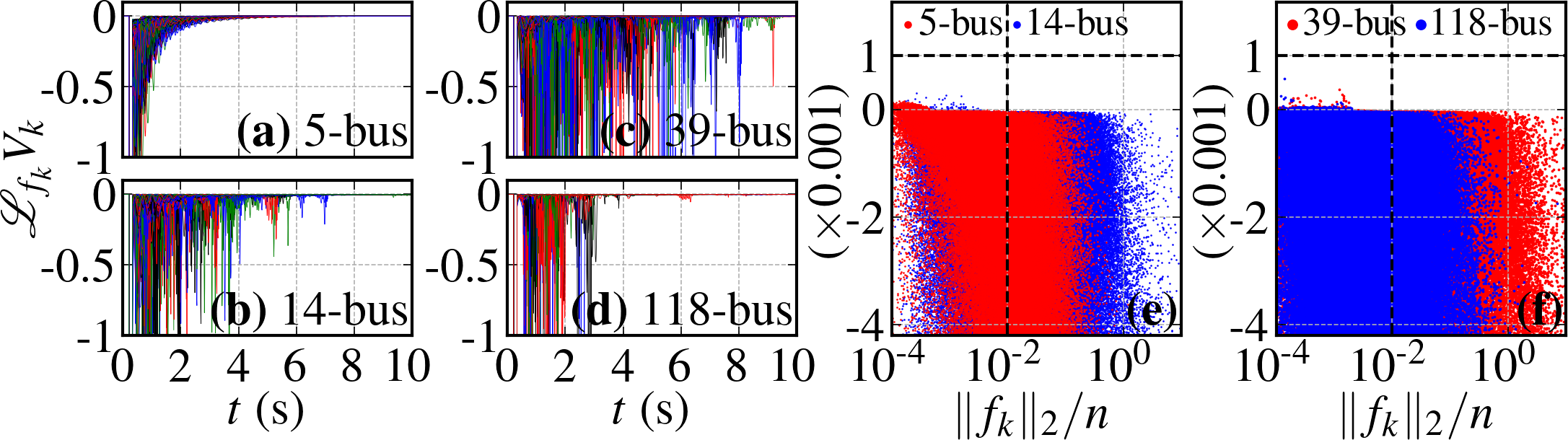}   
    \vspace{-10pt}
	\caption{(a)-(d) Time-domain curves of $\mathcal{L}_{f_k} V_k$ for the four test systems under various fault scenarios; (e)-(f) scatter plots of $\Vert f_k \Vert_2 / n$-$\mathcal{L}_{f_k}\! V_k$ after the faults. }
	\label{fig-8-2-r2}
    \vspace{-8pt}
\end{figure}

The validity of the obtained EF is further demonstrated via post-fault system state trajectories from 4 systems which are not used in training. 
These systems are the 5-, 14-, 39-, and 118-bus systems, with the same structure as the equal-sized IEEE test systems. 
For each system, 1000 different scenarios of $\bm{\rho}$ values taken from $\mathbb{P}_k$ and 3-phase short-circuit faults %(occurring at $t\!=\!0.1$ s and clearing after 0.1 s)
are used, in each of which the system is stable. 
Fig. \ref{fig-8-2-r2}-(a) to (d) show the time-domain curves of $\mathcal{L}_{f_k} V_k$ under these scenarios. 
Intuitively, the value of $\mathcal{L}_{f_k} V_k$ increases but remains non-positive after the faults, and converges to 0 as time progresses. 
This is consistent with the expected change of EF values after faults. 
Precisely, Fig. \ref{fig-8-2-r2}-(e) and (f) give the scatter plots of $\Vert f_k \Vert_2/n$-$\mathcal{L}_{f_k} V_k$ for all post-fault state points. 
It is seen that for these state points, $\mathcal{L}_{f_k} V_k \!\!<\!\! 0$ if $\Vert f_k \Vert_2/n \!\geq\! 0.01 \!\!=\!\! \phi$, and $\mathcal{L}_{f_k} V_k \!\!<\!\! 0.001 \!\!=\!\! \varepsilon$ otherwise. Thus the conditions for EFs are satisfied with the pre-set tolerance.

\subsubsection{Comparison of RoAs} 
\begin{figure}[t!]
	\centering 
    \includegraphics[scale=0.95]{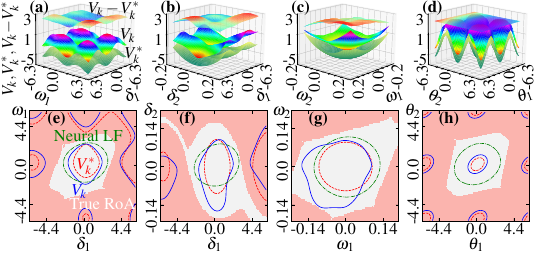}   
    \vspace{-10pt}
	\caption{(a)-(d) Surface plots of the neural-analytic EF at different cross-sections; (e)-(h) comparison of the estimated RoAs at different cross-sections. }
	\label{fig-8-2-r1}
    \vspace{-5pt}
\end{figure}

Taking the 5-bus system as an example, Fig. \ref{fig-8-2-r1}-(a) to (d) visualize the obtained neural-analytic EF $V_k$, the analytic EF $V_k^*$, and the NN component of $V_k$, i.e., $V_k \!-\! V_k^*$. 
It can be seen that at each cross-section, the overall shape of $V_k$ is similar to $V_k^*$, but their specific values are entirely different due to the NN component. 
Furthermore, Fig. \ref{fig-8-2-r1}-(e) to (h) compare the estimated RoAs obtained using the neural LF in \cite{4-1377}, $V_k^*$, and $V_k$, corresponding to the white regions enclosed by the green, red, and blue curves, respectively. 
It is observed that the neural-analytic EF $V_k$ yields a larger RoA than the analytic EF $V_k^*$ does. 
At the cross-sections of $\delta_1$-$\delta_2$ and $\omega_1$-$\omega_2$, the size of the RoA estimated using $V_k$ is close to that obtained using the neural LF, which although is specifically constructed for this 5-bus system. 
This observation indicates that the neural-analytic EF can provide less conservative stability assessment than the analytic EF.

\vspace{-4pt}
\section{Conclusion} 

This letter provides a learning-based paradigm for constructing EFs of power systems, expressed in neural-analytic form. 
The neural-analytic EF is more generalizable and scalable than the neural LF, and generates less conservative RoAs compared to the analytic EF. 
Future work will focus on higher-order system dynamic models and optimized training methods.

\ifCLASSOPTIONcaptionsoff
  \newpage
\fi

\vspace*{-4pt}
\bibliographystyle{IEEEtran}
\bibliography{4.bib}

\end{document}